\documentclass[12pt,preprint]{aastex}

\shorttitle{Albedo of the exoplanet HD 209458b}
\shortauthors{Rowe et al.}

\begin{document}

\title{The Very Low Albedo of an Extrasolar Planet:\\
MOST\footnote{MOST is a Canadian Space Agency mission, operated jointly by  Dynacon, Inc., and the Universities of Toronto and British Columbia, with assistance from the University of Vienna.} Spacebased Photometry of HD 209458}

\author{Jason F. Rowe}

\affil{NASA-Ames Research Park\\
MS-244-30\\
Moffett Field, CA 94035-1000\\
jasonfrowe@gmail.com}

\author{Jaymie M. Matthews}

\affil{University of British Columbia\\
6224 Agricultural Road\\ 
Vancouver BC, V6T 1Z1}

\author{Sara Seager}

\affil{Massachusetts Institute of Technology\\
Cambridge, MA 02159, USA }

\author{Eliza Miller-Ricci, Dimitar Sasselov}
\affil{Harvard-Smithsonian Center for Astrophysics \\ 
60 Garden Street, Cambridge, MA 02138, USA}

\author{Rainer Kuschnig}
\affil{Department of Physics and Astronomy, University of British Columbia \\ 
6224 Agricultural Road, Vancouver, BC V6T 1Z1, Canada}

\author{David B. Guenther}
\affil{Department of Astronomy and Physics, St. Mary's University\\
Halifax, NS B3H 3C3, Canada}

\author{Anthony F.J. Moffat}
\affil{D\'epartement de physique, Universit\'e de Montr\'eal \\ 
C.P.\ 6128, Succ.\ Centre-Ville, Montr\'eal, QC H3C 3J7, Canada}

\author{Slavek M. Rucinski}
\affil{David Dunlap Observatory, University of Toronto \\
P.O.~Box 360, Richmond Hill, ON L4C 4Y6, Canada}

\author{Gordon A.H. Walker} 
\affil{Department of Physics and Astronomy, University of British Columbia \\ 
6224 Agricultural Road, Vancouver, BC V6T 1Z1, Canada}

\author{Werner W. Weiss}
\affil{Institut f\"ur Astronomie, Universit\"at Wien \\ 
T\"urkenschanzstrasse 17, A--1180 Wien, Austria}

\begin{abstract}
Measuring the albedo of an extrasolar planet provides insights into its  atmospheric composition and its global thermal properties, including heat dissipation and weather patterns.  Such a measurement requires very precise photometry of a transiting system sampling fully many phases of the secondary eclipse.  Spacebased optical photometry of the transiting system HD 209458 from the MOST (Microvariablity and Oscillations of STars) satellite, spanning 14 and 44 days in 2004 and 2005 respectively, allows us to set a sensitive limit on the optical eclipse of the hot exosolar giant planet in this system.  Our best fit to the observations yields a flux ratio of the planet and star of 7 $\pm$ 9 ppm (parts per million), which corresponds to a geometric albedo through the MOST bandpass (400-700 nm) of $A_g$ = 0.038 $\pm$ 0.045.  This gives a 1$\sigma$ upper limit of 0.08 for the geometric albedo and a 3$\sigma$ upper limit of 0.17.  HD 209458b is significantly less reflective than Jupiter (for which $A_g$ would be about 0.5). This low geometric albedo rules out the presence of bright reflective clouds in this exoplanet's atmosphere.  We determine refined parameters for the star and exoplanet in the HD 209458 system based on a model fit to the MOST light curve.
\end{abstract}


\section{Introduction}\label{intro}

Among the roughly 250 planets discovered to date around nearby stars\footnote{The Extrasolar Planets Encyclopedia: http://www.exoplanet.eu}, about two dozen are observed to transit.  Multicolour optical photometry of the transits provides accurate measurements of the orbital inclination, and hence the mass and radius of the planet with respect to the parent star \citep[e.g.,][]{knu07}. Infrared measurements of the eclipses in thermal emission \citep{dem05a,cha05}  give brightness temperatures at specific wavelengths to constrain the atmospheric properties.  Measurement of the optical eclipse of such a system provides a geometric albedo, which when combined with the thermal emission data, sensitively probes the exoplanetary atmosphere to test for high-altitude clouds (such as silicate condensates) and possibly even weather patterns \citep{row06a}.

Models of the interior structure of giant exoplanets require appropriate boundary conditions -- planetary mass, radius and surface temperature -- which can be set by optical photometry of transiting hot extrasolar giant planets (EGPs).  HD 209458b, which orbits its host star at a distance of only 0.047 AU every 3.5 days, the thermal equilibrium temperature for the dayside photosphere can be estimated as 
\begin{equation}\label{eq:teq}
T_{eq}=T_* (R_*/2a)^{1/2} [f(1-A_B)]^{1/4},
\end{equation}
where the star and planet are considered blackbodies, $T_*$ and $R_*$ are the effective temperature and radius of the host star, the planet at distance $a$ with a Bond albedo of $A_B$ and $f$ is a proxy for atmospheric thermal circulation.  The Bond albedo, $A_B$, is the fraction of total power incident on a body scattered back into space. 

The geometric albedo $A_g$ is the ratio of brightness at zero phase angle compared to an idealized flat, fully reflecting and diffusively scattering disk.  From $A_g$, one can derive the ratio of the stellar and planetary fluxes 
\begin{equation}\label{eq:ag}
\frac{F_p}{F_*} = A_g \left(\frac{R_p}{a}\right)^2,
\end{equation}
where $R_p$ is the planetary radius and $a$, the separation between star and planet.   

If the composition of HD 209458b is similar to Jupiter, then the surface temperature of the exoplanet is the dominant variable driving the atmospheric chemistry, and hence, the albedo. This means that the albedo should be a function of both the star-planet separation, and the luminosity of the host star \citep{mar99}.  The relationship between albedo and surface temperature is a good tracer of chemical processes (such as the formation of molecules) in planetary atmospheres.
 
HD 209458 is a nearby (47 pc), bright ($V = 7.65$) Sun-like star (G0V) hosting a transiting hot EGP \citep[e.g.,][]{cod02}.  Spitzer observations at 24 $\mu m$ detected an eclipse at a depth of $0.26\% \pm 0.046\%$ which gives a brightness temperature (1130 $\pm$ 150 K) for the observed band \citep{dem05a}.  Estimating the equilibrium temperature is highly model-dependent \citep{sea05,for05,bar05,bur07b}.  Atmospheric models by \citet{for05} predict an effective temperature of 1442 K, assuming 4$\pi$ reradiation and solar metallicity. 

The brightness of the star and the short orbital period ($\sim$3.5 d) of the exoplanet made this planetary system an ideal candidate for photometry by the MOST (Microvariablity and Oscillations of STars) satellite.  The objectives of the observations were: (1) to measure accurately times of several consecutive transits to search for undetected Earth-mass planets in close orbits \citep{mil07}; (2) to search for transits of nearly-Earth-radius planets at other periods in the nearly continuous data sets \citep{cro07}; (3) and to measure or set a limit on the optical eclipse of the exoplanet HD 209458b.

The first results of our eclipse analysis \citep{row06a}, based on a MOST trial run of 14 days in August 2004 sampling 4 consecutive transits and 4 secondary eclipses, set upper limits on the eclipse depth of 53 ppm (parts per million) and on the geometric albedo of 0.25.  We present here improved measurements of the eclipse based on the original photometry and a second 44-day MOST run in 2005, sampling an additional 12 consecutive transits and eclipses.  The combined datasets give  1$\sigma$ upper limits on the eclipse depth of 16 ppm and on the geometric albedo of 0.08.

\section{MOST Photometry}

The MOST satellite \citep{wal03,mat04} houses a 15-cm optical telescope feeding a CCD photometer through a single custom broadband (400 - 700 nm) filter.  From its 820-km-high polar Sun-synchronous orbit, MOST can monitor stars in its equatorial Continuous Viewing Zone for up to 8 weeks without interruption.  

Photometry of HD 209458 was obtained in two runs, during 14 - 30 August 2004 and 1 August - 15 September 2005, for a total of 58 days of high-cadence photometry.  The data were collected in MOST's Direct Imaging mode \citep{row06a}, where a defocused image of the star is recorded in a subraster of the Science CCD, with 1.5-s exposures sampled every 10 s. Two fainter stars were simultaneously observed, HD 209346 (A2, V = 8.33) and BD +18 4914 (A0, V = 10.6) in the same way.  The latter was found to be a hybrid ($\delta$ Scuti + $\gamma$ Dor) pulsator \citep{row06b}.  We rejected  exposures with high cosmic ray fluxes which occur when MOST passes through the South Atlantic Anomaly (SAA), as well as data with background illumination values greater than 3000 ADU  due to scattered Earthshine modulated with the satellite orbital period (see \S \ref{errs}). This resulted in data sets with 106,752 and 334,245 points and duty cycles of 81\% and 89\% respectively for the 2004 and 2005 runs. 

\begin{figure}[ht]
\plotone{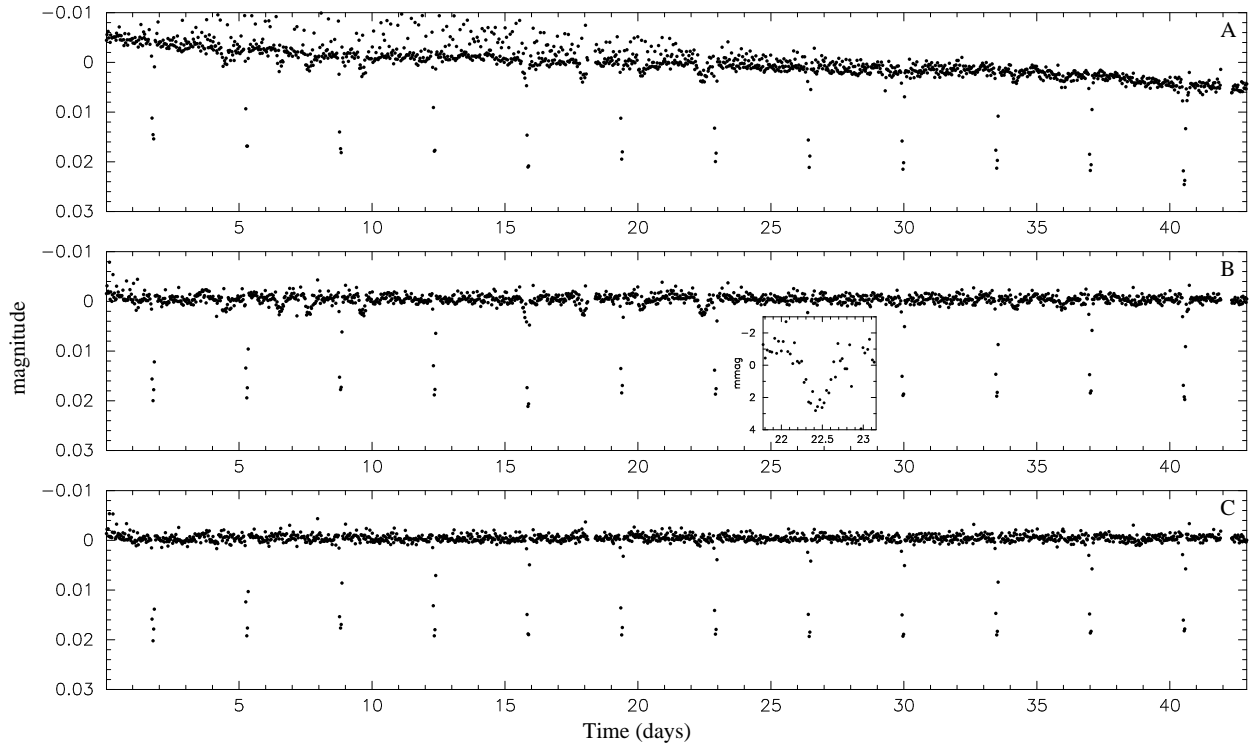}
\caption{The 2005 MOST photometry of HD 209458. Top: The unprocessed light curve. 
	Middle: The data after corrections for stray light correlations.  Bottom: The final 
    reduced light curve, after crosstalk corrections.  The data have been averaged in 
	40-min bins for clarity.  The large scatter in the first half of the light curve is because the stray light levels where comparitively much higher.}
\label{fig:datacor}
\end{figure}

The light curve of the 2004 run is shown in Figure 4 of \citet{row06a}, and the light curve of the 2005 run is plotted in Figure \ref{fig:datacor}.  This figure shows the raw photometry and the photometry after reduction as described by \citet{row06a}. The reduction procedures include the removal of the correlation between background light and instrumental stellar flux measurements, correction of low levels of crosstalk noise between the Science and Startracker CCDs (read out at different rates), and estimation of photometric errors as a function of the sky background level.  The last two steps (described in detail below) were critical to improve the limit on the optical albedo of the exoplanet and also for the transit timing \citep{mil07} and transit search \citep{cro07} conducted with this same set of photometry.

\subsection{Crosstalk Corrections}\label{ct}

The effects of crosstalk noise on the photometry can be seen in the upper and middle panels of Figure \ref{fig:datacor} as small dips ($\sim$ 2-4 mmag) in the light curve at times of 16, 18, 20 and 22.5 d.  Crosstalk (or ``video noise'') occurs when the CCD controller electronics interact with another electrical component due to incomplete grounding \citep{gil92}.  In the MOST instrument, the Science and Startracker CCDs are located close beside each other in the camera focal plane, and are electronically isolated from one another.  The aluminum structure of the spacecraft bus provides excellent grounding for the camera electronics, and crosstalk levels tested prior to launch were well below the original mission science requirements.  However, the transit and eclipse analyses of HD 209458 were not part of the original plan for MOST and these results are more sensitive to the intermittent crosstalk.  

Crosstalk happens when both the Science and Startracker CCDs are read out at the same time.  The duration of frame transfer and readout for the Science CCD is about 1 s, while the Startracker has a readout time of ${\sim}0.1$ s.  A noise band approximately 10 pixels wide appears on the Science CCD (corresponding to the ratio of readout times) which can overlap the subraster containing the target star.   This noise source does not obey Poisson statistics, which allows us to correct for it effectively. Without crosstalk, the noise for pixels used to estimate the sky background level in the subraster can be predicted from Poisson noise expected from the incident stray light plus read noise inherent to the detector electronics.  This can be compared directly to the standard deviation of the pixel intensity values.  Crosstalk introduces additional noise which changes the ratio of measured to predicted noise, which would be constant outside of crosstalk events.  The correlation of background light and instrumental magnitude (see  \S \ref{errs} below) does introduce changes in the noise ratio, but the time scale of this variation (the satellite orbital period of about 101.4 min) and the durations of crosstalk events (about 0.5 d) are so different that they are easily distinguished from each other.

The positive signal added by crosstalk noise will traverse through the downloaded Science CCD subraster as the times of the independent CCD readouts move out of synch. The background level will be overestimated, but the PSF (Point Spread Function) fit to the stellar image is largely unaffected.  Hence, the stellar flux is underestimated and the width of the crosstalk-affected pixel region moving across the CCD gives the characteristic {\it sawtooth} shape of the crosstalk artifacts seen in the upper two panels of Figure \ref{fig:datacor}. By correlating the background pixel scatter with the stellar photometry, the background level is rectified and the correct stellar flux value is recovered.  The bottom of Figure \ref{fig:datacor} shows the 2005 photometry after this correction.  In particular, notice the adjustment to the depth of the fifth transit during day16, which coincided with a crosstalk event.

\subsection{Photometric Error Estimation}\label{errs}

\begin{figure}[ht]
\plotone{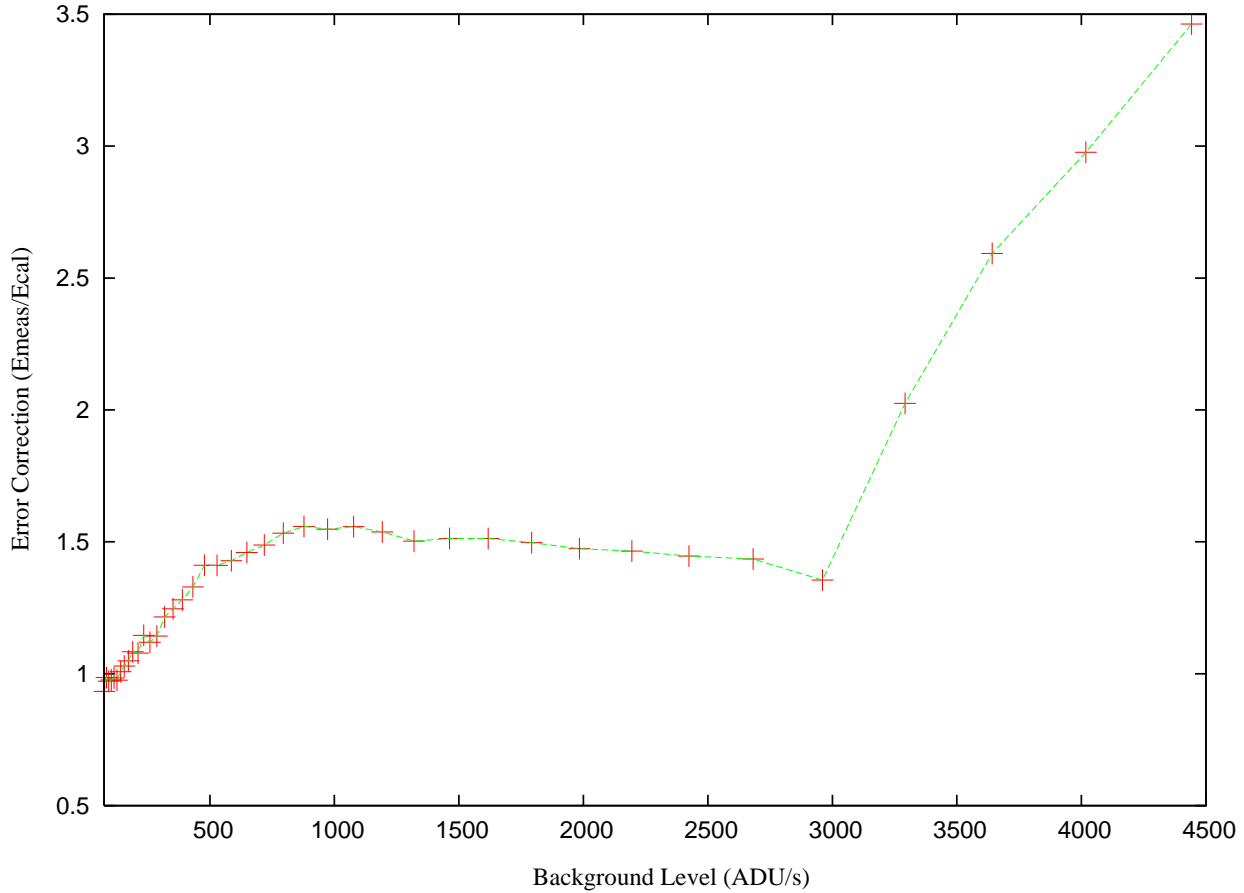}
\caption{Corrected photometric error estimates as a function of background level. 
	$E_{\rm meas}$ is the measured error based on the standard deviation of each bin 
	and $E_{\rm cal}$ is the expected error based on photon statistics.  The sharp 
	increase in the ratio of these two errors after 3000 ADU is due to saturation of the 
	detector.}
\label{fig:errcor}
\end{figure}

It is still unclear what causes the correlation of sky background to instrumental stellar magnitude readings in the data. To investigate this, we tested the validity of the predicted Poisson errors to the observed scatter, outside of times of crosstalk and times surrounding exoplanetary transits (about 5 hr every 3.5 d).  We binned the data by sky background value, with an equal number of data points for each bin.  We then calculated the expected scatter in each bin based on Poisson statistics and the measured standard deviation.  Figure \ref{fig:errcor} shows the ratio of predicted to measured scatter as a function of the background level.  At background levels of about 3000 ADU (Analog-to-Digital Units), the peak of the stellar PSF becomes saturated, since the 14-bit ADC (Analog-to-Digital Conversion) of the MOST CCD electronics has a saturation limit of 16 384 ADU.  This causes the large discrepancy between theory and measurement seen at this threshold in the figure.  At low background levels there is a linear increase up to about 1000 ADU, followed by a slower rising plateau until saturation. This change could be interpreted as a variation in CCD gain with signal.

We have corrected the error estimates with a linear interpolation over the binned sky background values for each photometric measurement.  This provides the correct weight for each measurement and the proper statistics for subsequent fitting to models of the light curve.

\subsection{The Phased Light Curve}

\begin{figure}[ht]
\epsscale{0.75}
\plotone{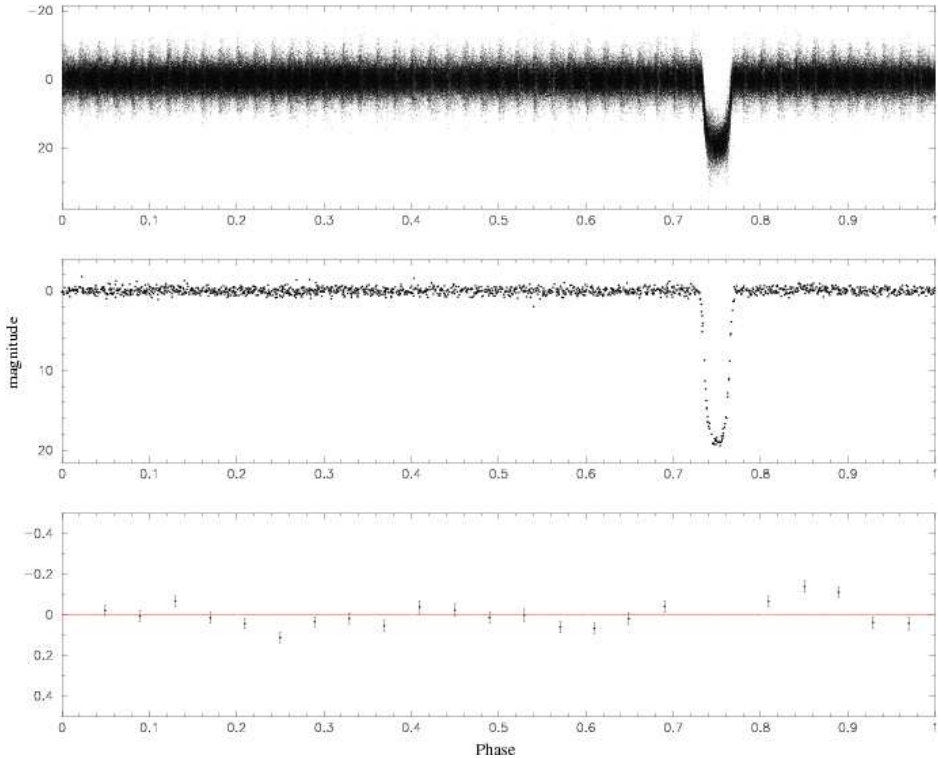}
\caption{The reduced 2004 and 2005 MOST photometry of HD 209458 phased to the period of the exoplanet.  Top: Unbinned data. The pattern outside of the phase of transit is due to the coincidental harmonic relationship (50:1) between the orbital frequencies of MOST and the exoplanet, so modulated scatter in the data due to scattered Earthshine is also in phase with the period of HD 209458b. Middle: The data averaged in 40-min bins.  Bottom: The data in bins of width 0.04 cycle in phase. Note the different magnitude scales for the three plots.}
\label{fig:data}
\end{figure}

Figure \ref{fig:data} presents  the 2004 and 2005 photometry plotted in phase with the orbital period of the exoplanet (see Table \ref{ta:fitpars}).  The top panel shows the unbinned data.  The repetitive pattern seen in these data is due to the increased photometric scatter during intervals of highest stray light modulated with the MOST orbital period of about 101.4 min. (This appears correlated to the planets orbits because, by coincidence, the orbital frequencies of the MOST satellite and HD 209458b are in a near-harmonic ratio of 50:1.)   The middle panel shows the data binned in 40-min intervals.  The effects of stray light are less obvious here, since we use weighted averages based on the photometric errors described above in \S\ref{errs}.   The planetary transit is at phase 0.75; the eclipse (not visible) at phase 0.25. The bottom panel shows the data averaged in bins of width 0.04 cycle in phase (${\sim}3.4$ hr) with $1\sigma$ error bars of about 25 ppm ($\mu$mag).  

\section{Modeling the Light Curve}

The light curve of a system with one transiting planet contains the following variations in phase with the exoplanet orbital period: (1) the transits themselves; (2) the eclipses; and (3) the changing flux from the planet as it goes through illumination phases during its orbit. There are also possible intrinsic variations in the star due to rotational modulation, some of which may be synchronised with the close-in planet's orbit (see, e.g., \citet{wal06}).

The Bond albedo can be written as,
\begin{equation}
A_B = A_g \ q
\end{equation}
where $A_g$ is the geometric albedo and $q$ is the phase integral.  From the definition of $A_g$ in Equation \ref{eq:ag} the geometric albedo must be observed at zero phase angle.  The chemical composition of the atmosphere, including particle size, can strongly influence the phase integral, such as strong backscatter \citep[for example, see][]{gre03}.  

First we place limits on the albedo based on comparison of observations adjacent to the eclipse.  The planetary transit lasts for $\sim$ 3.7 hours (and hence the eclipse).  To measure the secondary eclipse for the planet disappearing behind the star, we average the photometry over phase bins of width 0.044 ($\sim$3.7 hours) centred 0.5 phase away from the transit (at phase 0.25 in Figure \ref{fig:data}).  We then find the average of the photometry from two adjacent bins with the same width.  Our error is estimated by bootstrapping the means for each bin (see \S\ref{boot}).  This basic approach gives a flux ratio of the star and planet of $2.2 \times 10^{-5} \pm 2.9 \times 10^{-5}$. 

\begin{figure}[ht]
\plotone{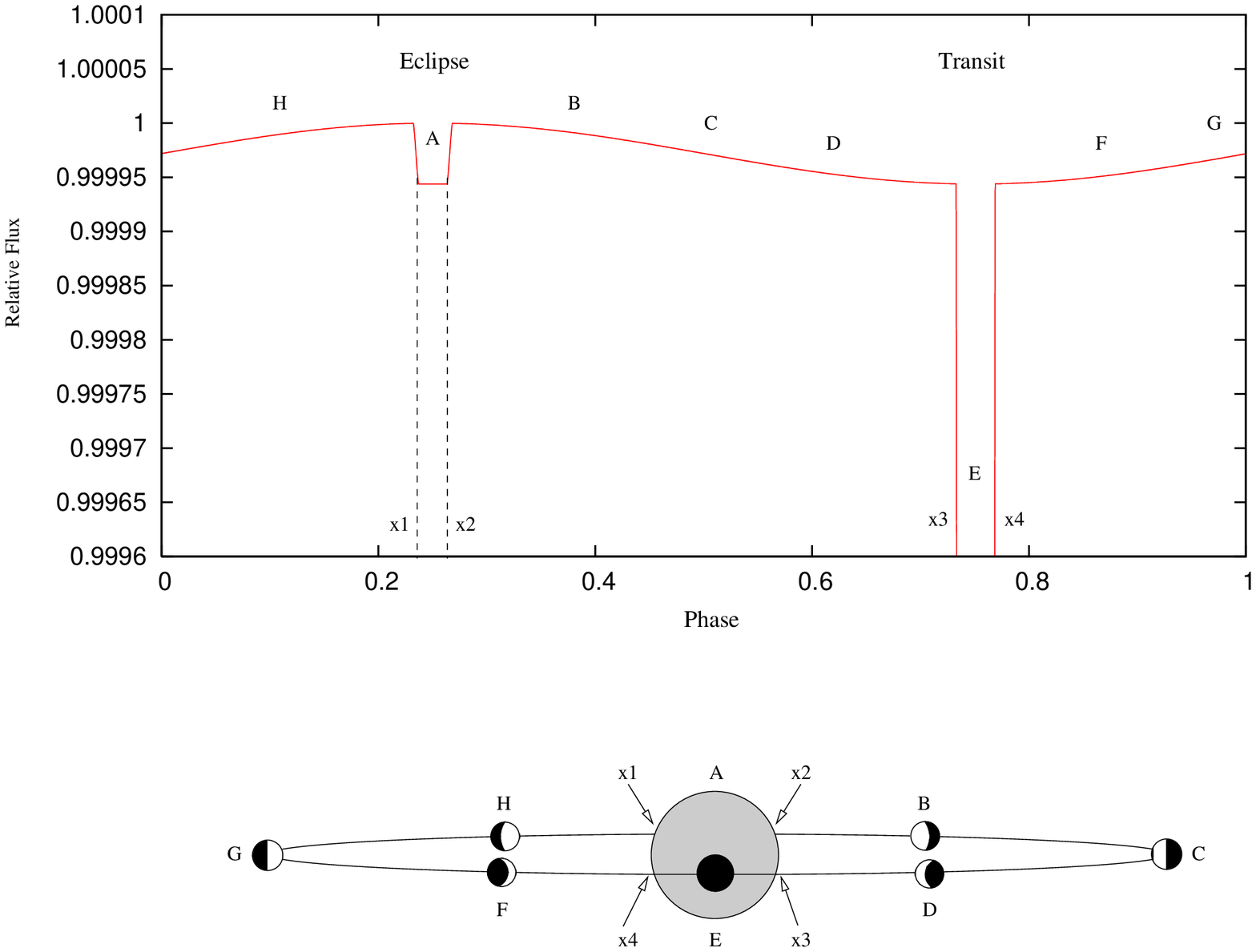}
\caption{A model of the flux changes during the orbit of HD 209458b.  The phase labeled A corresponds to the centre of the eclipse, when the flux is due to the star alone. Phase E corresponds to the centre of transit, where the normalized flux drops to about 0.98 (off the magnified scale of this plot). The sinusoidal curve (marked at phases B, C, D, F and H) is due to the changing illumination phase of the planet, as illustrated at the bottom of the figure.  Phases $x_1$ through $x_4$ indicate the ingresses and egresses of the eclipse and transit.}
\label{fig:eclipse}
\end{figure}

If we make the assumption that the planet scatters as a Lambert sphere then $q$ is simply a function of the phase of the planet from the observer's perspective.  For a planet in a circular orbit like HD 209458b, we have approximated the reflected light variation as a sinusoid (adequate for the phase variations from a Lambertian sphere for the current set of observations). Our function to model the light curve, including transits, is
\begin{equation}\label{model}
f = \frac{MA\left( R_*, R_p, a, i, c_n, t \right)  + \frac{F_p}{2 F_*} \left[ 1+ cos \left( \frac{2 \pi t}{P} + \phi \right) \right] sin(i) }{1 + \frac{F_p}{F_*} sin(i)},
\end{equation}
where MA is the normalized flux computed using the analytic expressions of \citet{man02} for a transiting planet, $F_*$, $F_p$, $ R_* $ and $ R_p $ are the fluxes and radii of the star and planet, $a$ and $P$ are the semi-major axis and period of the planet's orbit, $i$ is the inclination of the orbit relative to our line of sight, $c_n$ are the nonlinear limb darkening parameters for the star \citep{cla00} and $\phi$ is an arbitrary starting phase to coincide with the transit in the phase diagram.  A schematic of  this light curve model is shown in Figure \ref{fig:eclipse}.  The transit depth for HD 209458 extends to a relative flux of 0.98, far below the plotted scale.  The amplitude of the flux variations due to phase changes of the planet is equal to the depth of the eclipse of the planet by the star.  

\begin{figure}[ht]
\epsscale{0.75}
\plotone{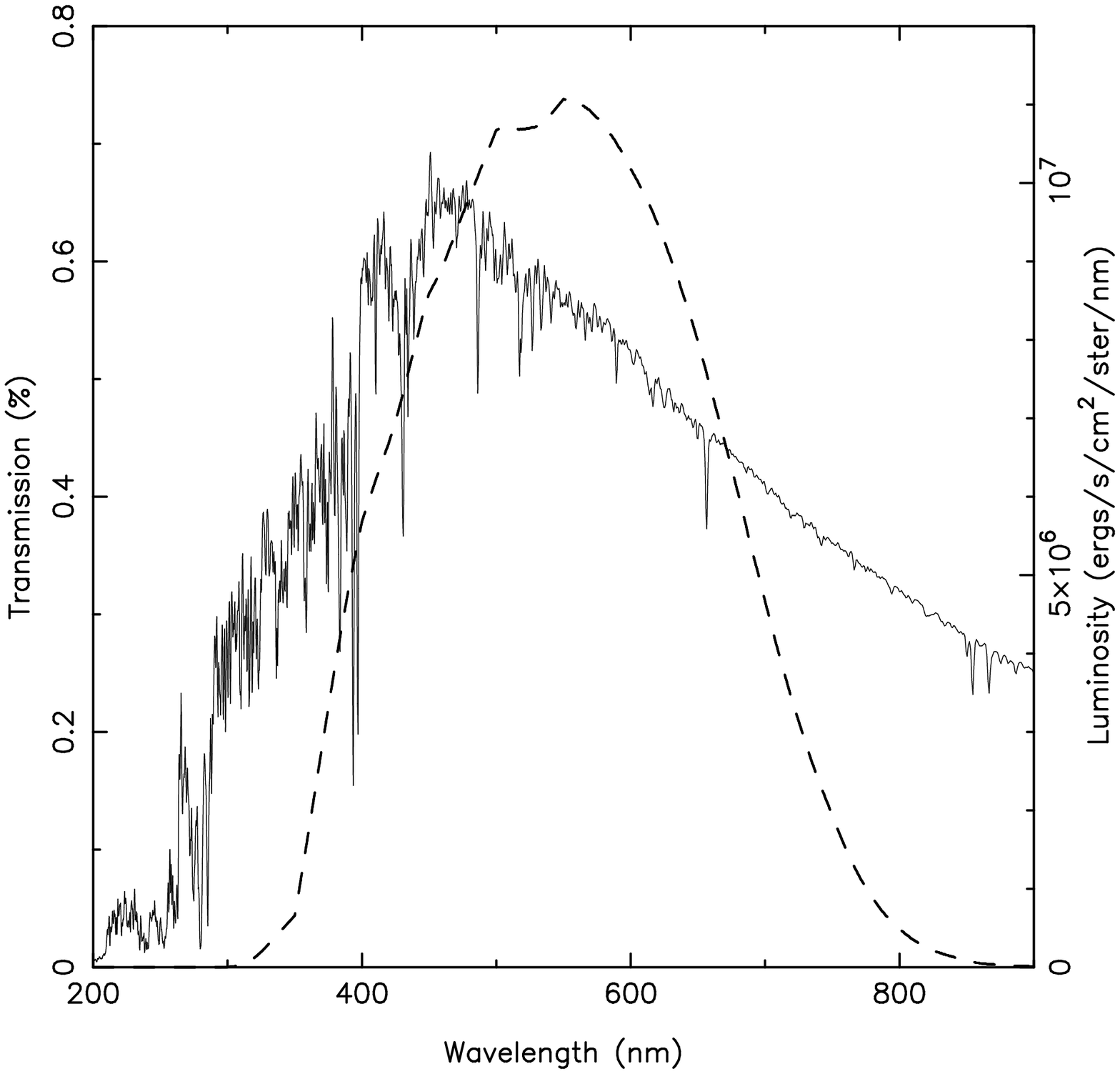}
\caption{Total system throughput for the MOST satellite optics and CCD detector is shown as the dashed line.  A Kurucz model with T$_{\rm eff}$ = 6100 K and log g = 4.38 cgs, representitive of HD 209458a is also plotted.}
\label{fig:trans}
\end{figure}

We adapt a Bayesian approach to our best-model-fit minimization so that we can incorporate priors in our fits.  The probability function that we wish to maximize is
\begin{equation}\label{fiteqn}
p(y_1,...,y_n) = \frac{p(y_1|I)...p(y_n|I)p(D|y_1,...,y_n,I)}{p(D,I)},
\end{equation}
where $y_1$ to $y_n$ are the 13 model parameters listed in Table \ref{ta:fitpars}, $p(y_1|I)$ to $p(y_n|I)$ are the corresponding priors, $p(D|y_1,...,y_n,I)$ is the likelihood function and $p(D,I)$ is a normalization factor.  Having priors is important as single-band photometry gives few constraints on the orbital inclination of the exoplanet due to degeneracy of this parameter with the radii of the planet and star. Our adapted priors for the orbital inclination, orbital period, mass and radius of the planet are from \citet{knu07}, based on multi-band HST photometry with limb-darkening information to estimate the inclination angle.  We take our prior for the mass of the planet from \citet{lau05}.  Limb darkening parameters for a non-linear model were determined from a Kurucz model representative of HD 209458 show in Figure \ref{fig:trans}, kindly provided by \citet{knu07}, scaled according to the MOST custom passband.  For Sun-like host stars, the peak of the emitted spectral energy distribution is in the optical range, so the MOST bandpass geometric albedo is a good approximation to the mean integrated value of the geometric albedo. 

Priors are given by
\begin{equation}\label{prior}
p(y_i)=\frac{1}{\sqrt{2\pi}e_i}{\rm Exp}\left( \frac{-(y_i-\mu_i)^2}{2e_i^2}\right),
\end{equation}
where $\mu_i$ is the prior value of $y_i$, with a $1\sigma$ uncertainty of $e_i$, as listed in Table \ref{ta:fitpars}.  We do not fit explicitly for the limb darkening parameters.  In the case of fixed parameters, $p(y_i) = \delta(y_i-\mu_i)$ and $p(y_i) = 1$ when no prior information is available.  The likelihood function is given by
\begin{eqnarray}\label{likelihood}
p(D|y_1,...,y_n,I) & = & {\rm Exp}[-\chi^2]\
\prod_{i=1}^n\frac{1}{\sqrt{2\pi}\sigma_i},\\ 
\chi^2          & = & \sum_{i=1}^n \frac{(d_i - f(x_i;y_1,...,y_n))^2}{\sigma_i^2},
\end{eqnarray} 
where $\sigma_i$ is the photometric uncertainty as calculated in \S\ref{errs}.

\begin{figure}[ht]
\plotone{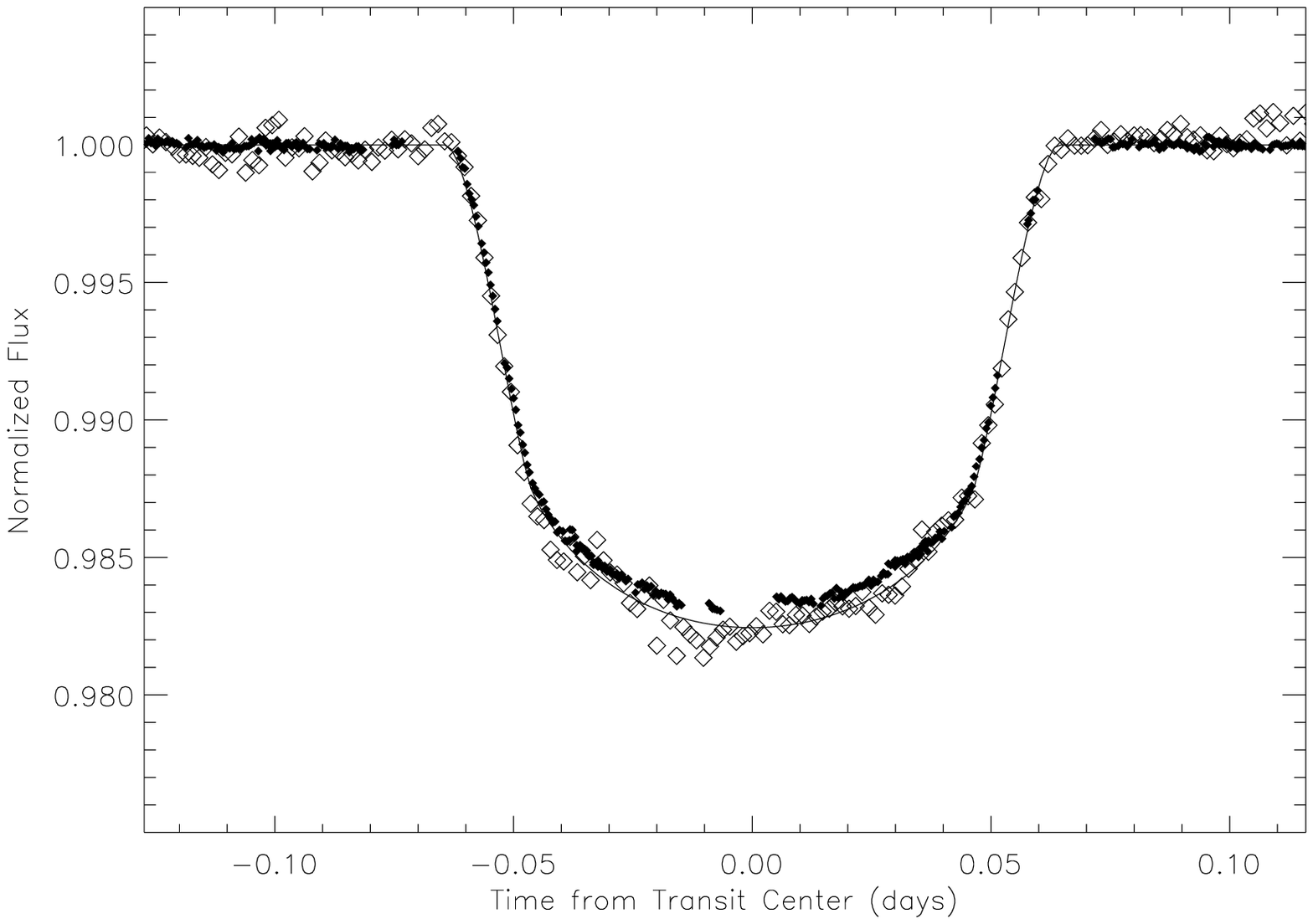}
\caption{Comparison of HST and MOST photometry of the transit of HD 209458b. The MOST data are open diamonds and the HST data (scaled to the MOST bandpass) as dots.  The curve is our best-fitted model for the MOST light curve.}
\label{fig:transithst}
\end{figure}

We find the maximum of equation \ref{fiteqn} using a downhill simplex model based on the Amoeba routine in Numerical Recipes \citep{pre92}.  We did not adopt any prior for the radius of the planet, as a direct comparison of MOST and HST photometry of the HD 209458 transits, scaled to the same passband, produces different depths of transit.   Using the spectroscopic observations of \citet{knu07} and the MOST bandpass, HST photometry was rescaled to match MOST photometry.  Figure \ref{fig:transithst} compares the averaged transit for the two sets of photometry. The MOST data indicate a deeper transit than the HST data.  This may indicate a systematic error in the measured amplitude of thr transit from MOST data.  It may also be related to the non-differential photometry for HST.  Unlike MOST, HST cannot continuously observe HD 209458.  Instead the complete transit much be pieced together from different orbits as the HST field-of-view will typically be occulted by the Earth during some portion of the orbit.  This causes a thermal change in the optical telescope assembly which in turn changes the optical focus, producing the well documented {\it breathing effect} which is relfected in the photometry \citep{mak06}. Keeping the planetary radius as a free parameter, our best-fitted value for the model is given in column 3 of Table \ref{ta:fitpars}.

\subsection{Bootstrap Error Analysis}\label{boot}

To estimate uncertainties in our fitted parameters, we use a bootstrap technique.  This involves randomly selecting data from the original time series and generating a new time series, with replacement.  Replacement means that any data point can be chosen more than once, but the total number of points is always the same, so some points will not be included in the new datasets.  Each generated data set has a noise profile similar to the original time series and repeating the fitting procedure on a series of these randomised data sets produces a robust error distribution for our best-fitted parameter values.  For other examples and discussion of the bootstrap method, we refer the reader to \citet{cam06} and references therein.

We performed ${\sim}20,000$ bootstrap iterations.  Figure \ref{fig:fitplots} shows the bootstrap results for some key parameters in our model fit.  Table \ref{ta:fitpars} gives $1\sigma$ errors for all fitted parameters.  These uncertainties were estimated by assuming a normal distribution and calculating the standard deviation of the bootstrap sample.    

\section{An Upper Limit on the Albedo of HD 209458B}

Our best fit to the flux ratio of the planet and star ($F_p/F_*$) is $7 \times 10^{-6}$ with a $1\sigma$ upper limit of $1.6 \times 10^{-5}$.  Applying the best-fitted value to equation \ref{eq:ag} gives the geometric albedo measured through the MOST filter: $A_{\rm MOST} = 3.8 \pm 4.5\%$.  

\begin{figure}[ht]
\epsscale{0.30}
\plotone{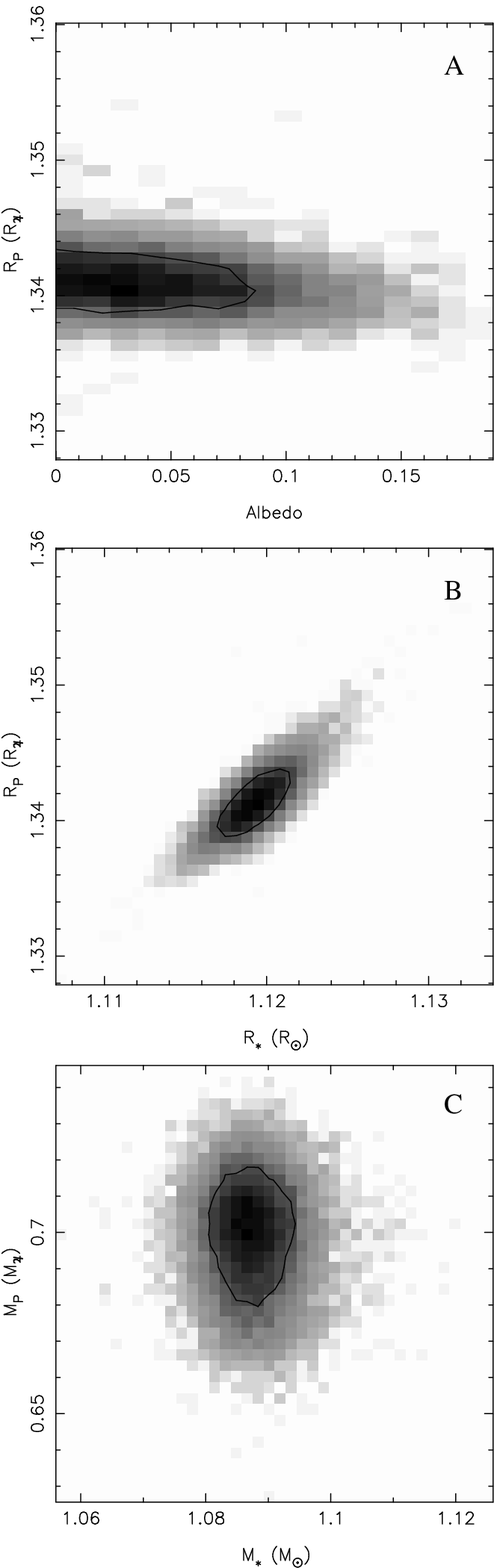}
\caption{Bootstrap results for HD 209458 parameter fits to MOST photometry. Contours outline the 68\% confidence region.  Top: Planet radius vs. geometric albedo through the MOST passband. Middle: Constraints on the radii of the star and planet. Bottom: Constraints on the masses of the star and planet.}
\label{fig:fitplots}
\end{figure}

The top panel of Figure \ref{fig:fitplots} shows the bootstrap error analysis for $A_{\rm MOST}$ as a function of the planet radius $R_p$.   The use of priors heavily constrains the system parameters for the planet (such as radius and mass) that are allowed.  A glance at the errors stated in Table \ref{ta:fitpars} reveals that these errors are unrealistically small.  Therefore, we also tested our fits by removing the prior information.  The derived errors (also listed in Table \ref{ta:fitpars}) are more in tune with other studies such as that of \citet{knu07}. Photometry alone gives no information about the mass of the planet or star and our unconstrained fits that give uncertainities on the mass of the star are larger than the mass of planet from spectroscopy.  The main subject in this paper is the flux ratio of reflected light from the planet compared to the host star and most of the other parameters are from fitting the shape of the transit, which applies to only a small fraction of the data.  The flux ratio of the planet and star is largely independent of the transit fitting parameters. 

In the MOST bandpass (shown in Figure \ref{fig:trans}) the Solar System giant planet geometric albedos are all greater than 0.4 and have a Bond albedo ($A_B$) greater than geometric albedo observed over all wavelengths ($A_{g{\rm TOT}}$),  If we assume that $A_g \sim A_{g{\rm TOT}}$ for HD 209458b and that $0.67 < A_{g{\rm TOT}}/A_B < 1$ based on arguments presented in \citet{row06a} we can estimate $A_B < 0.12$ at the $1 \sigma$ confidence level.  

Rayleigh scattering, Mie scattering and molecular absorption are the dominant mechanisms that determine the reflected and emitted spectra of an EGP \citep{mar06}.  In the bluer portion of the optical spectra (wavelengths shorter than 600nm) Rayleigh scattering in a clear atmosphere will reflect a large fraction of the stellar flux outwards. In the red portion (wavelengths greater than 600nm) photons will be absorbed deep in the atmosphere that will make the reflected spectrum relatively dark.  The strong incident UV flux can produce a rich mixture of compounds from molecules producing a haze that can absorb incident UV photons and will darken the appearance of the planet in the blue part of the spectrum \citep{mar99}.  Jupiter and Saturn, which receive much less UV flux than HD~209458, contain unidentified species at the level of a few parts per $10^{10}$ that decrease the blue and green geometric albedos by a factor of 2 \citep{kar94,bur07}.  While equilibrium photochemistry in HD~209458b cannot produce long-chain hydrocarbon hazes \citep{lia04} owing to the high temperatures, the importance of non-equilibrium photochemistry, or photochemistry of S and N, compounds has not yet been explored.

In the absence of clouds, all hot Jupiter models do predict extremely low visible-wavelength geometric albedos, due to strong, broad absorption lines of neutral atomic Na and K. These atoms are known to suppress the emitted visible-wavelength flux of brown dwarfs at similar atmospheric temperatures \citep{lie01}. Some model atmospheres include self-consistent cloud formation in a 1D, complete cloud cover scenario (\citet{ack01}, \citet{coo03}). Nevertheless, the physics of cloud formation is a process that is not well understood or constrained both for particle size and densities which are governed by the competing effects of condensation and coagulation versus sedimentation (Marley et al. 1999).  The low reflected fraction of incident radiation readily rules out reflective clouds.  New Spitzer observations at 3.6, 4.5, 5.8 and 8.0 $\mu$m \citep{knu07b} indicate a temperture inversion which requires an extra, unknown absorber at low pressures \citep{bur07b}.  Our low albedo limit means that, if the absorber is a cloud, then it must not be highly reflective.

\begin{figure}[ht]
\epsscale{0.75}
\plotone{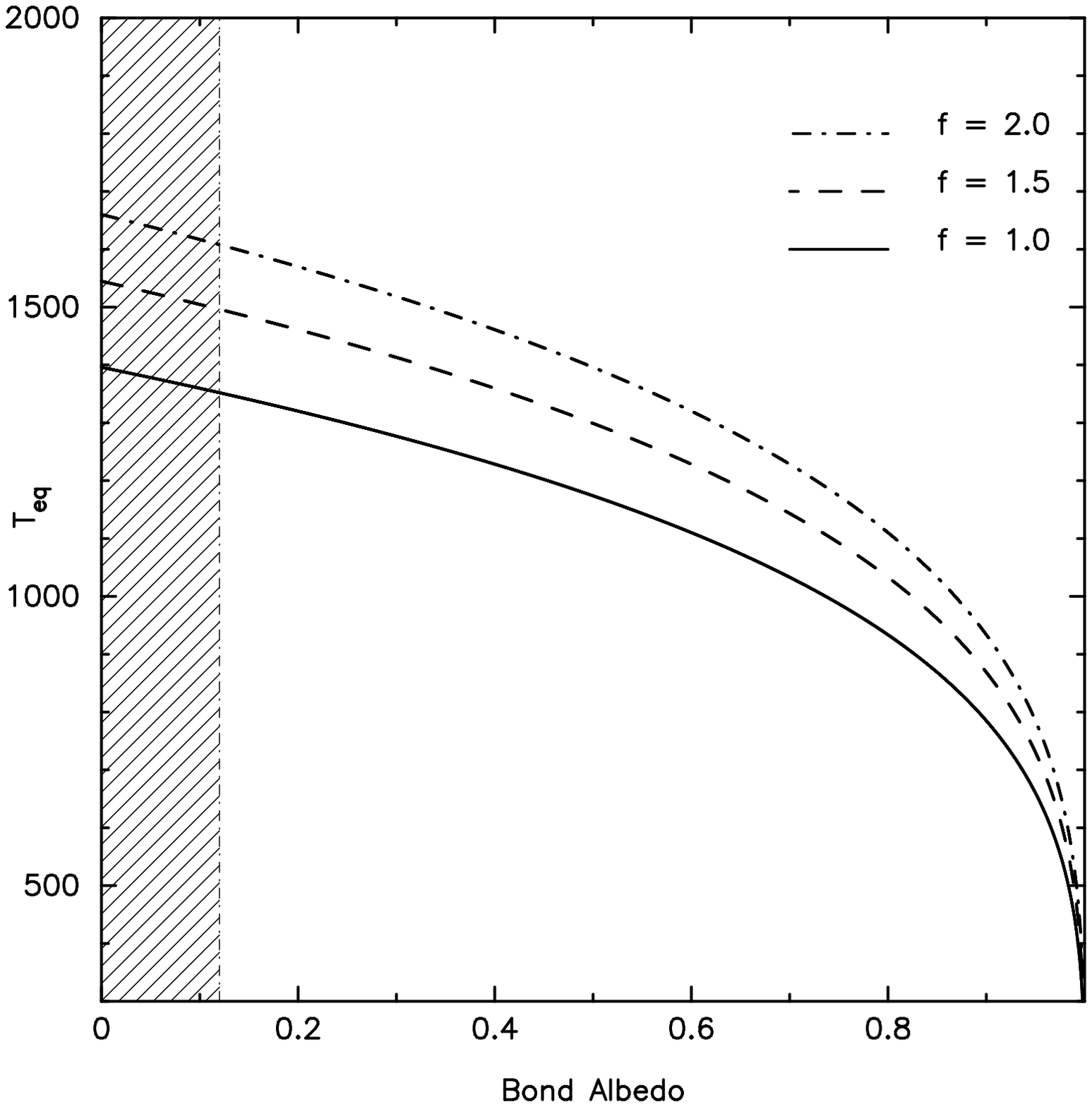}
\caption{The dayside $T_{eq}$ for HD209458b as a function of $A_B$ for different values of $f$ (see equation~\ref{eq:teq}). The 1$\sigma$ limit of $A_B$ is shown as a vertical dashed line.}
\label{fig:TAb}
\end{figure}

The efficiency of the planetary atmosphere to transport heat from the day to night side is commonly parametrized by the surface area over which the planet reradiates absorbed stellar flux.  For a fully mixed isotropic atmosphere, the planet will reradiate over 4$\pi$ steradians, whereas, without circulation the planet only reradiates over 2$\pi$ steradians.  HD 209458b is expected to be tidally locked.  Infrared measurements at 8 $\mu$m by \citet{cow07} place 2 $\sigma$ limits on the variations in IR flux with phase at 0.0015.  Our low Bond albedo measurement requires that planet distribute at least 35\% (at the 1 $\sigma$ limit) of the absorbed stellar energy on the night side of the planet.  This means that $f$ from Equation \ref{eq:teq} must be less than 2.  This agrees with \citet{bur07b} who find models with circulation can give reasonable fits to Spitzer measurements.  The atmospheric depth at which the bulk of the stellar radiation is absorbed determines whether advection or radiative transfer dominates energy transport and the efficiency with which a planet circulates heat to the nightside of the planet \citep{sea05}. Figure \ref{fig:TAb} plots the Bond albedo versus the equilibrium temperature for different values of $f$.  The horizontal line shows the 24 $\mu$m brightness temperature which gives a lower limit on $T_{eq}$ and the verticle line shows the 1 $\sigma$ upper limit on $A_B$.  This constrains the equilibrium temperature to be $1400 K < T_{eq} < 1650 K$ depending on the efficiency of thermal ciruclation in the planet's atmosphere.

\section{Conclusions and the Future}

HD 209458b is less reflective than Solar System giant planets such as Jupiter.  Our measurements place a 1$\sigma$ upper limit on $F_p/F_*$ of $1.6 \times 10^{-5}$. The inferred low Bond albedo $<0.12$ rules out presence of highly reflective clouds in the atmosphere of HD 209458b and is also consistent with non-cloudy atmospheric models.  It also constrains the equilibrium temperature to be between $1400$ and $1650K$   The largest unknown in determining $T_{eq}$ is the efficiency with which the planet distributes heat from the dayside to nightside of the planet.  Further infrared measurements at different phases of the planet orbit will place stronger constrains on this parameter.

The MOST satellite re-observed the HD 209458 system in August - September 2007 for approximately 4 weeks.  A new observing mode has been added that allows images to be stacked onboard the satellite before downloading.  This dramatically increases the number of photon counts that can be recorded.  Instead of obtaining a 1.5-second exposure every 10 seconds one can obtain 1 exposure every 15 seconds composed of 15 stacked 1.5-second exposures.  This has decreased scatter in the data by a factor of $\sim$3.  MOST will either detect the secondary eclipse or place a significance limit of a few percent on the albedo.  Analysis of these data is underway.  

\acknowledgments

The contributions of JMM, DBG, AFJM, SR, and GAHW are supported by funding from the Natural Sciences and Engineering Research Council (NSERC) Canada.  RK is funded by the Canadian Space Agency. WWW received financial support from the Austrian Science Promotion Agency (FFG - MOST) and the Austrian Science Fonds (FWF - P17580)

\begin{deluxetable}{ccccc}
\tablecolumns{6}
\tablewidth{0pc}
\tablecaption{Fitted Parameters for HD 209458 Lightcurve.\label{ta:fitpars}}
\tablehead{
\colhead{$y_i$} & \colhead{Prior} & \colhead{Best fit} & \colhead{Errors} & \colhead{Units} \\
\colhead{} &  \colhead{$(u_i \pm e_i)$} & \colhead{with Priors} & \colhead {without Priors} & \colhead{}
}
\startdata
$M_*$ &  $1.101 \pm 0.064$ & $1.083 \pm 0.005$ & $0.1$ & $M_\sun$ \\
$M_P$ & $0.69 \pm 0.05$ & $0.69 \pm 0.01$ & $1.0$ & $M_j$ \\
$R_*$ & $1.125 \pm 0.02$ & $1.118 \pm 0.002$ & $0.03$ & $R_\sun$ \\
$R_P$ & -- & $1.339 \pm 0.002$ & $0.04$ & $R_j$ \\
$P$ & $ 3.52474859  \pm 3.8\times10^{-7}$ & $3.5247489 \pm 2\times10^{-7} $ & $1\times10^{-6}$&days \\
$i$ & $ 86.929 \pm 0.01 $ &  $86.937 \pm 0.003$ & $0.2$ & deg \\
$A_g$ & -- & $0.038 \pm 0.045$ & $0.050$ & \\
$\phi$ & -- & $-1.57206 \pm 0.0001$ & $0.0002$ & rad \\
$zpt$ & -- & $-0.00001 \pm 1\times10^{-5}$ & $1\times10^{-5}$ & mag \\
$c1$ & fixed & 0.410769 & & \\
$c2$ & fixed & -0.108929& & \\
$c3$ & fixed &  0.904020& & \\
$c4$ & fixed & -0.437364 & & \\
\enddata
\caption{Table of parameters to describe the MOST HD 209458 light curve.  Column (1) lists the 13 parameters used to describe the light curve.  The priors in column (2) refer to constraints used to derive the best fit parameters listed in column (3). Column (4) gives the errors on the fitted parameters if no priors are used.  Column (5) gives the units for each row in the table.}
\end{deluxetable}


\begin{thebibliography}{otherstuff}

\bibitem[Ackerman \& Marley(2001)]{ack01} Ackerman, A.S., Marley, M.S. 2001, \apj, 556, 872
\bibitem[Barman et al.(2005)]{bar05} Barman, T.S., Hauschildt, P.H., Allard, F. 2005, \apj, 632, 1132
\bibitem[Burrows et al.(2005)]{bur05} Burrows, A., Hubeny, I., Sudarsky, D. 2005, \apjl, 625, L135
\bibitem[Burrows et al.(2007a)]{bur07} Burrows, A., Hubeny, I., Budaj, J., Hubbard, W.B. 2007, \apj, 661, 502
\bibitem[Burrows et al.(2007b)]{bur07b} Burrows, A., Hubeny, I., Budaj, J., Knutson, H., Charbonneau, D., 2007, astro-ph/07093980
\bibitem[Cameron et al.(2006)]{cam06} Cameron, C., Matthews, J.M., Rowe, J.F., Kuschnig, R. Guenther, D.B., Moffat, A.F.J., Rucinski, S.M., Sasselov, D., Walker, G.A.H. Weiss, W.W. 2006, CoAST, 148, 57
\bibitem[Charbonneau et al.(2005)]{cha05} Charbonneau, D., Allen, L.E., Megeath, S.T., Torres, G., Alonso, R., Brown, T.M., Gilliland, R.L., Latham, D.~W., Mandushev, G., O'Donovan, F.T., Sozzetti, A. 2005, \apj, 626, 523
\bibitem[Claret(2000)]{cla00} Claret, A. 2000 \aap, 363, 1081
\bibitem[Cody \& Sasselov(2002)]{cod02} Cody, A.M., Sasselov, D. 2002, \apj, 569, 451
\bibitem[Cooper et al.(2003)]{coo03} Cooper, C.S., Sudarsky, D., Milsom, J.A., Lunine, J.I., Burrows, A. 2003, ApJ, 586, 1320
\bibitem[Cowan et al.(2007)]{cow07} Cowan, N.B., Agol, E., Charbonneau, C. 2007, \mnras, 379, 641 
\bibitem[Croll et al.(2007)]{cro07} Croll, B. et al. 2006, \apj, 658, 1328
\bibitem[Deming et al.(2005a)]{dem05a} Deming, D., Seager, S., Richardson, L.~J., \& Harrington, J.\ 2005, \nat, 434, 740 
\bibitem[Fortney et al.(2005)]{for05} Fortney, J.J., Marley, M.S., Lodders, K., Saumon, D., Freedman, R. 2005, \apj, 627, L69
\bibitem[Fortney et al.(2007)]{for07} Fortney, J.J., Marley, M.S., Barnes, J.W. 2007, \apj, 659, 1661
\bibitem[Gilliland(1992)]{gil92} Gilliland, R.L. 1992, in ASP Conf. Ser. 23, Astronomical CCD Observing and Reduction Techniques, ed. S. B. Howell (San Francisco: ASP), 68
\bibitem[Green et al.(2003)]{gre03} Green, D., Matthews, J., Seager, S., Kuschnig, R. 2003, \apj, 597, 590
\bibitem[Liebert(2001)]{lie01} Liebert, J. 2001, Ultracool Dwarfs: New Spectral Types L and T. Edited by Hugh R. A. Jones and Iain A. Steele. Berlin Heidelberg: Springer, 2001, p.3
\bibitem[Karkoschka(1994)]{kar94} Karkoshka, E. 1994, Icarus, 111, 174
\bibitem[Knutson et al.(2007a)]{knu07} Knutson, H.A., Charbonneau, D., Noyes, R.W., Brown, T.M., Gilliland, R.L. 2007, \apj, 655, 564
\bibitem[Knutson et al.(2007b)]{knu07b} Knutson, H.A., Charbonneau, D., Allen, L.E., Burrows, A., Megeath, S.T., 2007, astro-ph/07093984
\bibitem[Laughlin et al.(2005)]{lau05} Laughlin G., Marcy, G.W., Vogt, S.S., Fischer, D.A., Butler, R.P. 2005, \apj, 629, L121
\bibitem[Liang et al.(2004)]{lia04} Liang, M.-C., Seager, S., Parkinson, C.D., Lee, A.Y.-T., Yung, Y.L. 2004, \apj, 605, 61
\bibitem[Makidon et al.(2006)]{mak06} Makidon, R.B., Lallo, M.D., Casertano, S., Gilliland, R.L., Sirianni, M., Krist, J.E. 2007, in: D.R Silva \& R.E. Doxsey, Observatory Operation: Strategies, Processes, and Systems, (eds.) 62701L
\bibitem[Mandel \& Agol(2002)]{man02} Mandel, K., Agol, E. 2002, \apj, 580, 171
\bibitem[Marley et al.(1999)]{mar99} Marley M.S., Gelino, C., Stephens, D., Lunine, J.I., Freedman, R 1999, \apj, 513, 879
\bibitem[Marley et al.(2006)]{mar06} Marley M.S., Fortney, J., Seager, S., Barman, T. 2006 astro-ph/0602468
\bibitem[Matthews et al.(2004)]{mat04} Matthews, J.M., Kuschnig, R., Guenther, D.B., Walker, G.A.H., Moffat, A.F.J., Rucinski, S.M., Sasselov, D., Weiss, W.W. 2004, \nat, 430, 51 
\bibitem[Miller-Ricci et al.(2007)]{mil07} Miller-Ricci, E., Rowe. J.F., Sasselov, D., Matthews, J.M., Guenther, D.B., Kuschnig, R., Moffat, A.F.J., Rucinski, S.M., Walker, G.A.H., Weiss, W.W. 2007, \apj, submitted
\bibitem[Press et al.(1992)]{pre92} Press, W.H. Teukolsky, S.A. Vetterling, W.T. Flannery, B.P. 1992, Numerical Recipes in Fortran 77 Second Edition, Cambridge University Press, 678
\bibitem[Rowe et al.(2006a)]{row06a} Rowe J.F., Matthews, J.M., Seager, S., Kuschnig, R., Guenther, D.B., Moffat, A.F.J., Rucinski, S.M., Sasselov, D., Walker, G.A.H., Weiss, W.W. 2006a, \apj, 646, 1241
\bibitem[Rowe et al.(2006b)]{row06b} Rowe J.F., Matthews, J.M. Cameron, C., Bohlender, D.A., King, H., Kuschnig, R., Guenther, D.B., Moffat, A.F.J., Rucinski, S.M., Sasselov, D., Walker, G.A.H., Weiss, W.W. 2006b, CoAST, 148, 34
\bibitem[Seager \& Sasselov(1998)]{sea98} Seager, S., Sasselov, D. 1998, \apj, 502, L157
\bibitem[Seager et al.(2000)]{sea00} Seager, S., Whitney, B.A., Sasselov, D.D. 2000, \apj, 540, 504
\bibitem[Seager et al.(2005)]{sea05} Seager, S., Richardson, L.~J., Hansen, B.~M.~S., Menou, K., Cho, J.~Y-K., \& Deming, D.\ 2005, \apj, in press
\bibitem[Sudarsky et al.(2000)]{sud00} Sudarsky, D., Burrows, A., Pinto, P. 2000, \apj, 538, 885
\bibitem[Sudarsky et al.(2005)]{sud05} Sudarsky, D., Burrows, A., Hubeny, I. Li, A. 2005, \apj, 627, 520
\bibitem[Walker, Matthews et al.(2003)]{wal03} Walker, G., Matthews, J.M., Kuschnig, R., Johnson, R., Rucinski, S. Pazder, J., Burley, G., Walker, A., Skaret, K., Zee, R., Grocott, S. Carroll, K., Sinclair, P., Sturgeon, D., Harron, J. 2003, \pasp, 115, 1023 
\bibitem[Walker et al.(2006)]{wal06} Walker, G.A.H., Matthews, J.M., Ruschnig, R., Rowe, J.F., Guenther, D., Moffat, A.F.J., Rucinski, S., Sasselov, D., Seager, S., Shkolnik, E., Weiss, W. 2006, in: L. Arnold, F. Bouchy, \& C.Moutou, Tenth Anniversary of 51 Peg-b: Status of and prospects for hot Jupiter studies, (eds.) 267 
\end{thebibliography}
\end{document}